\documentstyle {article}
\begin{document}
%
\def \sp#1#2{ \la #1 \v #2\ra}  
\def\lr #1{\mathrel{#1\kern-.75em\raise1.75ex\hbox{$\leftrightarrow$}}}
\def\larr #1{\mathrel{#1\kern-.75em\raise1.75ex\hbox{$\leftarrow$}}}
\def\rarr #1{\mathrel{#1\kern-.75em\raise1.75ex\hbox{$\rightarrow$}}}
\def \vf {\varphi}
\def \o {\omega}
\def \v {\vert}
\def \a {\alpha}
\def \b {\beta}
\def \s {\sigma}
\def \p {\psi}
\def \ra {\rangle}
\def \la {\langle}
\def \tp {t_0+\epsilon}
\def \tm {t_0-\epsilon}
\def\PRL{ {\sl Phys. Rev. Lett.}   }
\def\PR { {\sl Phys. Rev.  }       }
\def\NP { {\sl Nucl. Phys.}        }
\def\PL { {\sl Phys. Lett.}        }
\def\CMP { {\sl Comm. Math. Phys. }     }
\def\MPL { {\sl Mod. Phys. Lett.}        }
\overfullrule=0pt
\def\sqr#1#2{{\vcenter{\vbox{\hrule height.#2pt
          \hbox{\vrule width.#2pt height#1pt \kern#1pt
           \vrule width.#2pt}
           \hrule height.#2pt}}}}
\def\square{\mathchoice\sqr68\sqr68\sqr{4.2}6\sqr{3}6}
\def\lrpartial{\mathrel{\partial\kern-.75em\raise1.75ex\hbox{$\leftrightarro
w$}}}
\begin{flushright}
ULB-TH 93/16\\
UMH-MG 93/03\\
October 1993\\
\end{flushright}
\vskip 2.5 truecm
\centerline{\bf{Quantum Back Reaction on a Classical Field}}
\vskip 1. truecm
\centerline{R. Brout\footnote{e-mail: smassar @
ulb.ac.be},\addtocounter{footnote}{-1}
S. Massar\footnotemark \ \footnote{Boursier IISN},
S. Popescu\footnote{e-mail: spopescu @ ulb.ac.be}}
\centerline{Service de Physique Th\'eorique, Universit\'e Libre de
Bruxelles,}
\centerline{Campus Plaine, C.P. 225, Bd du Triomphe, B-1050 Brussels,
Belgium}
\vskip 5 truemm
\centerline{R. Parentani\footnote{e-mail: renaud @ vms.huji.ac.il}}
\centerline{Departement of Theoretical Physics,
The Racah Institute of Physics,}
\centerline{The Hebrew University of Jerusalem, Givat Ram Campus,
Jerusalem 91904, Israel}
\vskip 5 truemm
\centerline{Ph. Spindel\footnote{e-mail: spindel @ umh.ac.be}}
\centerline{M\'ecanique  et Gravitation, Universit\'e de Mons-Hainaut,
Facult\'e des
Sciences,}
\centerline{15 avenue Maistriau, B-7000 Mons, Belgium}
\vskip 1.5 truecm
{\bf Abstract }We show how to apply post selection in the context of weak
measurement of Aharonov and collaborators to construct the quantum back
reaction on a classical field. The particular case which we study in this
paper is pair creation in an external electric field and the back reaction
is
the counter field produced by the pair \underline {as} it is made. The
construction leads to a complex electric field obtained from non diagonal
matrix elements
of the current operator, the
interpretation of which is clear in terms of weak measurement. The
analogous construction applied to
black hole physics (thereby leading to a complex metric) is relegated to a
future paper.
\vfill
\newpage

\section{\bf Introduction}

The problem of the back reaction on a classical field  due to a quantum
process is to a great extent an unsolved problem in modern physics.
Ignorance of the solution is (at the least) one of the barriers towards the
solution of the so-called unitarity problem posed by black hole physics
 \cite{hawk1} ,
i.e how does the black hole evaporation engendered by
gravitational collapse convey all the information that has gone into
the initial conditions describing the quantum state prior to the
collapse?

The state of the art of black hole reaction
physics is at present rather primitive. For example in the black hole
problem one calculates the expectation  value  of the energy-momentum due
to
the Hawking radiation \cite{hawk2} \ (for a review see ref. \cite{birreld}
), i.e. the thermal
radiation that occurs at assymptotically large Schwarzschild times where
the
collapsing body hugs the horizon infinitely  closely. This expectation
value then serves as a source to the classical Einstein's equations.
Clearly the fluid--like
description characteristic of this semiclassical approach begs the question of
the details of how the gravitational field, better that part the
wave function which describes gravity, reacts to the quantum
process, i.e. tunnelling \cite{pbt}, which results in the emission of a
single (or a few)
Schwarzschild quantum(a). At best the semiclassical theory describes some
coarse-grained mean evolution. So it is very difficult to imagine that such
an approach to the unitary issue can reveal the information, phases among
other things, which is being sought.

The considerations of this paper are conceived as a (very small) step
towards the solution of this problem. We study the back reaction on the
electric field due to production of pairs induced by its presence. As
pointed out by several authors \cite{muller,stephens,bps,pb}, this problem,
owing to the presence of horizons (caused by the constant acceleration
induced
by the field on charged quanta) bears many analogies to black hole
evaporation. Indeed, the amplitudes for production in the two cases have
essentially the same mathematical structure. But there are profound
differences which cause the black hole problem to be much more difficult.
For
example the ``member" of the pair that is not measured (or measurable!) by
the
Schwarzshild observer is hidden from him by the horizon (except if it turns
out in the end that the complete back reaction kills the horizon; but then
the
problem is more complicated yet). Of course in the electric case both
members
of the produced pair are accessible to observation and subtle problems of
loss
of information do not arise (provided one allows for ubiquitous
measurement). Nevertheless interesting non trivial problems occur even in
this case when one delves into the quantum mechanics of the production of a
single pair.

The problem of how to deal with electroproduction in the mean has been
dealt
with by Cooper et al. \cite{mottola} in a semi-classical approach similar
to that brought to bear in black hole physics. Their approach is valid in
the
presence of large density of pairs. Our purpose here is to ask more
detailed
questions appropriate to the opposite case where pairs are rare. How does
one formulate with precision the question: ``\/what is the field caused by
the
creation of a single pair?\/" We recall that a pair is created by tunneling
\cite{casher,stephens,bps} out of a region of spatial dimension $\v\Delta
x\v$ , which is of order
$a^{-1}$ where $a$ is the acceleration ($=E/m$), $E$ being the electric
field with
the charge of the field quanta absorbed into its definition, and $m$ their
mass.
For  $\v\Delta x\v> a^{-1}$ the produced quanta propagate on mass shell in
opposite directions and one expects in a wave packet description that this
asymptotic region can be adequately described in conventional terms. But
what happens \underline{as} the pair is created in the tunneling region?
After all
$\nabla . E=\rho$ is a constraint, i.e fluctuations of the quantum
observable $\rho$ give rise to fluctuations of (the longitudinal part of)
$E$ which is constrained to follow them. So a detector of the $E$ field
will
be sensitive to the existence of the pair in this quantum region.

In our effort to answer this question we found difficulty for the following
reason. If in the distant past one has vacuum then in the future, pairs are
produced everywhere and all the time. Since the final state is
\underline{not} a
single pair state we must address the problem of how to isolate the field
$E$
produced by a single pair.  The problem is further complicated when one
introduces modes \cite{bps,pb}, the bases of the second quantized field. In
mode
language a  single particle mode in the distant past gives rise to a single
particle in the future plus a pair. To extract the field due to this
latter, it
is necessary to subtract the effect of that part of the mode which is due
to
the single particle. One can think of all kind of recipies, such as
substracting the field due to a WKB function, but we, at least, never
succeeded along these lines in going beyond just that - invention of \`a priori
recipes, not intrinsic to the problem and inevitably approximate in
character. In fact we were stymied by the problem for a long time because
the
only ``natural" construction for $\rho$ which cut out the contribution from
the single particle portion of a mode was to take the product of the
incoming particle and antiparticle wavefunctions (see Appendix 2 for the
formal definitions of these words). The construction, however, gave rise to
a
complex current. Taking the real part, was once more, merely a recipe.
Obviously, in the present case the issue is purely academic (as well as the
problem), but in the corresponding black hole problem, the unitarity issue
must be handled with ``kid gloves" and one is motivated to look for a
rigorous prescription that is intrinsic to the problem. The problem of
electric production is sufficently simple to allow for a rigorous analysis.

The measurement of $E$ we envision is a conditional measurement: we suppose
that we start with vacuum and that at a later time a single pair is
produced and
detected. The solution to our problem lies in the ``pre and post selection"
formalism developed by  Y. Aharonov and collaborators \cite{aharo},
 which is devised
to handle such conditional measurements. Being operational, in that it
delivers
information on how measuring devices react, it delivers a satisfactory
response
to our problem, free of any contamination resulting from a priori
substraction
recipes. In particular, the ``weak" measurements considered by Aharonov et
al. involve
 non diagonal matrix elements of the
operator $\rho$ and hence complex values of $E$, and gives them an
operational
interpretation. It should also be mentioned that the post selection
formalism
allows to answer the question: what is $E$ in the region where the pair
tunnels out? In fact it is only in this region that $E$ is complex.

{}From the conventional point of view of transitions in quantum mechanics
what we are doing is, so to speak, ``getting inside the golden rule". A
rate formula is obtained in evaluating transition probabilities by
waiting sufficiently long so that the amplitudes which destructively
interfere no longer contribute, thereby leaving only those pieces that
build up the long time behavior, usually characterised by energy
conservation (Poincar\'e's secular perturbations). In our problem, or the
black hole problem, the characterization is in terms of ``on-mass shell"
propagation, i.e. the asymptotic propagation of a solution of the
homogeneous wave equation in the presence of a given external field which
represents a single particle. This asymptotic solution occurs in our case
after tunneling, a period of ${\cal O}(a^{-1})$. Within this period
oscillations occur
because the virtual excitations are in the process of materialization.
There is no sense
in talking about the rate of production of a given pair until after this
tunneling time
elapses. In this sense we are asking for information outside the context of
the golden rule,
i.e. before the rate concept emerges. See the end of Section 3 for details.

Much of the paper is concerned with technical matters which rely on
previously published material. To keep the paper self contained, relevant
previous work is presented in Appendices. To keep the reading smooth some
new technical material is also written up in
appendices. Section 2 contains the analysis of particle production in terms
of minimal wave packets. Section 3 gives the analysis of the matrix
elements
of postselection in terms of these wave packets. The imaginary part of the
matrix element of $\rho$ will be seen to arrise from the tunneling region
only. Its measurement is analyzed in operational terms.

\section {\bf Wave Packets}

By virtue of the problem we have set for ourselves the construction of wave
packets
 will be essential
to our  analysis. In Appendix 1 we
have summarized second quantization in terms of modes, solutions of the
Klein-Gordon equation in the presence of a constant external electric
field.
This is carried out in two gauges: $A_0=-Ex$, $A_x=0$ and $A_0=0$, $A_x=Et$
(called spatial and temporal gauges respectively). For definiteness we use
the
latter to construct wave packets. It goes without saying that the packets
are gauge independent up to an irrelevant phase.

In terms of the basis functions of eq. (\ref{apov}) we make the packet
construction
to describe an in-particle
\begin{equation}
\psi^{in}_{k_0 x_0, p} (x,t) =
\int dk f(k - k_0)  e^{-ik x_0} \vf^{in}_{k , p}(x,t).\label{ai}
\end{equation}
The wave packet is centered at $x=x_0$ and $t_0=k_0/E$ (since $\vf^{in}_{k
,
p}(x,t)$ is a function of $t+k/E$, and the quadratic potential of the
Schr\"odinger problem is centered at $t=k$). Without loss of generality we
take a wave packet centered at the origin, i.e. $x_0=0$, $k_0=0$, and
henceforth we drop the indices $x_0$, $k_0$.

The notion of a minimal wave packet emerges from the asymptotic behavior of
the modes
$\vf^{in}_{k,p}$ as $ t \rightarrow \pm \infty$ where the WKB approximation
prevails
\begin{eqnarray}
&\lim_{t \rightarrow - \infty} \vf^{in}_{k,p} &\simeq
e^{ikx} e^{i E (t+k/E)^2 /2}  \nonumber\\
&\lim_{t \rightarrow + \infty} \vf^{in}_{k,p} &\simeq
\a^* e^{ikx} e^{-i E (t+k/E)^2 /2} - \b e^{ikx} e^{i E (t+k/E)^2
/2}\label{aii}
\end{eqnarray}
where we have kept only the leading exponential behavior of
$\vf$ and $\a$, $\b$ are the Bogoljubov coefficients eq. (\ref{apoix}). The
analysis is
greatly facilitated by taking $f$ to be gaussian, whereupon in the
asymptotic
regions the packet (always taking $x_0=0$, $k_0=0$) is
\begin{eqnarray}
& \lim_{t \rightarrow - \infty} \psi^{in}_p
&\simeq
\int dk\ e^{-k^2/2\s^2} e^{i k x} e^{i E (t+k/E)^2 /2}   \nonumber\\
& &\simeq e^{i E t^2/2} e^{-(x+t)^2/2 \Sigma^{ 2}_-}   \nonumber\\
& & \nonumber\\
&\lim_{t \rightarrow + \infty} \psi^{in}_p
&\simeq
\int dk e^{-k^2/2\s^2} ( \a^*  e^{i k x}e^{-i E (t+k/E)^2 /2} -
\b  e^{i k x} e^{i E (t+k/E)^2 /2})   \nonumber\\
& &\simeq \a^* e^{-i E t^2/2} e^{-(x-t)^2/2 \Sigma^{2}_+}
- \b e^{i E t^2/2} e^{-(x+t)^2/2 \Sigma^{2}_-}\label{aiii}
\end{eqnarray}
where
\begin{equation}
\Sigma^{2}_+ = \left( {1 \over \s^2} + {i\over E}\right)
\qquad \Sigma^{2}_- = \left( {1 \over \s^2} - {i\over E}\right)\label{aiv}
\end{equation}

The asymptotic width of the wave packets is given by
$[ {\rm Re} (1 / \Sigma_\pm^2) ]^{-1/2}$. If
$Im(\s) \neq 0$, then the two branches of $\psi^{in}_{p}$ as $t \rightarrow
+ \infty$
have unequal widths. This implies an asymmetric treatement of the particles
and antiparticles. As nothing warrants such partiality we take $Im(\s)=0$.
In this case the asymptotic width of the wave packets is
$[ {\rm Re} (1 / \Sigma_+^2) ]^{-1/2}=
[ {\rm Re} (1 / \Sigma_-^2) ]^{-1/2} = [(E^2 + \s^4) / E^2 \s^2]^{1/2}$ and
is minimized  by the choice $\s^2=\s^2_{min}=E=ma$ corresponding to wave
packets of width $[ {\rm Re} (1 / \Sigma_{min}^2) ]^{-1/2}= [ {2
/E}]^{1/2}$.

We now introduce an integral representation of the Whittaker functions,
linear combinations of eigenfunctions of the product of operators $UV$
where $U = p-
x /\sqrt {2}$ and $V = p+x /\sqrt {2}$  as explained in \cite{bal,pb} in
physical terms.
It is also this integral representation which serves as a basic tool
in revealing interesting propreties of these functions. It now has the
added
lustre that  the integral over $k$ in eq. (\ref{ai}) is gaussian
thereby leading to a simple integral representation for
gaussian packets which is nothing more than another Whittaker function.
This
we now display, where we work in the temporal gauge and choose for
simplicity units
where $E=1$.

The integral representation of $\vf_{k, p}^{in} (x,t)$ we use is (with $\mu
^2 \equiv m^2/2E$)
\begin{eqnarray}
&\vf_{k, p}^{in} (x,t) &= { e^{-\pi \mu^2/4} \over (2)^{1/4} }
e^{i k x}
D_{i \mu^2-1/2} \left [ e^{3 i \pi/4} \sqrt{2}
(t+ k)\right]\nonumber \\
& &={ e^{-\pi \mu^2/4} \over (2 )^{1/4} \Gamma(1/2-i\mu^2)}
e^{i k x} e^{i(k+t)^2/2} \nonumber \\
& & \qquad \times \int_0^{+\infty} du\ e^{(1-i)(k+t)u - u^2/2}
u^{-i\mu^2 - 1/2}\label{av}
\end{eqnarray}
whereupon the gaussian wave packet centered at $t_0=0$ and $x_0=0$ is
\begin{eqnarray}
 &\psi^{in}_{p}(x,t) &={ e^{-\pi \mu^2/4} \over (2 )^{1/4}
\Gamma(1/2-i\mu^2)}\!
\int_{- \infty}^{+ \infty}\!\! dk
{e^{-{k^2 \over 2 \s^2}} \over \sqrt{2 \pi} \s}\nonumber \\
& &\qquad \times \int_0^{+\infty}\!\! du\
e^{i k x +i{ (k+t)^2 \over 2} + (1-i)(k+t)u - {u^2 \over 2 }}
u^{-i\mu^2 - 1/2}   \nonumber\\
& &= { e^{-\pi \mu^2/4} \over (2 )^{1/4}}{1 \over \Sigma \s}
\lambda^{1/2 - i
\mu^2} e^{i t^2/2} e^{-{ (t+x)^2\over 2 \Sigma^2}} e^{z^2/4}
D_{i \mu^2-1/2}(z)\label{avi}
\end{eqnarray}
where $\Sigma^2 = (1/\s^2 -i)$, $\lambda^2 = (1-i\s^2)/(1+i\s^2)$,
$z= -\lambda (1-i) (t+ i\s^{2} x)/(1-i\s^2)$
with $\v {\rm arg}\ \Sigma \v < \pi /4$ and
$\v {\rm arg}\ \lambda \v < \pi /2$. The calculation leading to the second
equality is
carried out in detail in
Appendix 2.

The three other interesting wave packets centered
at $k=0$ and $x=0$ are
\begin{eqnarray}
&\psi^{in}_a (x,t) &=  \psi^{in}_p (-x,t)   \nonumber\\
&\psi^{out}_p (x,t) &=  \psi^{in*}_p (x,-t)   \nonumber\\
&\psi^{out}_a (x,t) &=  \psi^{in*}_p (-x,-t) \ .\label{avii}
\end{eqnarray}
Since the Bogoljubov coefficients eq. (\ref{apoix}) are independent of $k$,
the
functions $\psi^{in(out)}_{p(a)}$ are related  one to another
by the same Bogoljubov transformation. Note that this simplification will
not
occur when considering wave packets in the analogue black hole problem.

These
packets will be used in Section 3 to build the localized back reaction
arising
from their production. Figs Ia--c and II show how they carry current in
well
localized regions of space-time.

\section {\bf The Back Reaction Due To A Single Pair}

In Appendix 3 we have summarized the theory of a
weak measurement carried out on a post selected state.
The relevant matrix element of the operator $A$ which is
coupled to the detector is ${< p\v A \v i> \over
<p\v i>}$ where $\v i \ra$ is the initial state of the
system and $\v p \ra$ is the post selected state. Applied
to our problem, $\v p \ra$ is a single pair and $\v i \ra$ is vacuum. For
example one might
envision an electron which is sent through the region of production of
this pair. The measurement of the back reaction due to the pair emitted
is then the change of its path induced by the back
reaction $\Delta E$ whose source is the charge density,
$(\rho)$, due to the pair. Since $\Delta E$ in one dimension is obtained
by integrating the constraint $\nabla . \Delta E= \rho$, the study of back
reaction
is then reduced to the evaluation of the matrix element
\begin{equation}
{\ _{out}\! \la 1_p 1_a \v J_\mu \v 0\ra _{in}
\over  \ _{out}\! \la 1_p 1_a \v 0 \ra_{in} }\label{bi}
\end{equation}
where  $J_\mu$ is the current operator.
Here $\v 0\ra_{in}$ ($\v 0\ra_{out}$) means vacuum with
respect to in (out) modes and  $\v 1_p 1_a \ra_{out}$ is the
postselected state consisting of a pair obtained from the tensorial product
of
one--particle states such as
\begin{equation}
\v 1_p\ra_{out} =\int \! dk f(k) a^{out +}_k \v 0\ra_{out}\label{bii}
\end{equation}
 where $f$ is the same function as in
eq. (\ref{ai}) and similarily for the antiparticle.

It will be seen in the sequel that the evaluation of eq. (\ref{bi})
leads to some background noise related to zero point
fluctuations in addition to the
piece that is relevant to calculate physical quantities that arises
from the creation of the specific pair. In preparation for the
subtraction of such irrelevancies it is meet first to calculate matrix
elements of $\phi^*(x) \phi (x^\prime)$ between states of
interest. Matrix elements of operators constructed from
bilinear forms of $\phi$ such as the current $j_\mu$ or
the energy momentum tensor may then be obtained by
differentiation and going to the coincidence limit. In
the present case we then first caculate
$\ _{out}\! \la 1_p 1_a \v
\phi^*(x) \phi (x^\prime) \v 0\ra_{in}$ for which one
finds the equality (Appendix 4)
\begin{equation}
{ \ _{out}\! \la 1_p 1_a \v
\phi^*(x) \phi (x^\prime) \v 0\ra
_{in} \over
\ _{out}\! \la 1_p 1_a \v 0\ra_{in} }
=
-{ 1 \over \a \b^*} \psi^{in *}_{p}(x)
\psi^{in *}_{a}(x^\prime)
+ G_F(x,x^\prime)\label{biii}
\end{equation}
where $G_F$ is the familiar in-out propagator, encoding the
abovementioned noise
\begin{equation}
G_F(x,x^\prime) = { \ _{out}\! \la 0 \v
\phi^*(x) \phi (x^\prime) \v 0\ra_{in} \over
\ _{out}\! \la 0 \v 0\ra_{in} }\label{biv}
\end{equation}

The first term of eq. (\ref{biii}) refers to the postselected pair
with $\psi^{in}_{p}$ and $\psi^{in}_{a}$ being the
associated particle and antiparticle wavefunctions
respectively. The second term is background noise which results from
the structure of $\v 0\ra_{in}$ which encodes the
potentiality of making all the other pairs ( recall
$\v 0\ra_{in}$ is a linear combination of out states).
This noise term is independent of what the postselected
state is i.e. of the quantum numbers  of the pair, of
the number of pairs, etc... Furthermore it is
translation invariant. Therefore it can be isolated by measuring it once.
For
example if the post selected state is out-vacuum  only the noise
term appears and can be measured in isolation.

The above procedure is fine for isolating the
contribution of the first term. Nevertheless the noise
term is formally infinite -therefore to be handled
carefully. Furthermore it is of interest in its own
right, in that it contains a physically meaningful finite part.
Therefore we have written an extensive Appendix (Appendix 5) devoted to
this
point and in particular to how we adapt the Hadamard subtraction scheme to
our
problem. We find that the current carried by the noise term vanishes (as it
should: if the post selected state is out-vacuum, no pairs are created and
the current vanishes). For other operators quadratic in $\phi$ (the
energy momentum, etc ...) it is non vanishing.

Returning to our main theme, we then have in hand the
relevant matrix element obtained (after subtracting the noise) by
operating on eq. (\ref{biii}) with ${\cal D}_\mu$ as follows
\begin{eqnarray}
&j_\mu (x) &= { \ _{out}\! \la 1_p 1_a \v
J_\mu (x) \v 0\ra_{in} \over
\ _{out}\! \la 1_p 1_a \v 0\ra_{in} }   \nonumber\\
& & \nonumber \\
& &= -{1 \over \a^* \b}
( \psi^{in *}_a  (-i{ \cal D}_\mu^*  \psi^{in *}_p ) +
\psi^{in *}_p  (i{ \cal D}_\mu  \psi^{in *}_a )) \ . \label{bv}
\end{eqnarray}
Asymptotically $j_\mu (x)$ behaves as it should, classicaly. It vanishes in
the past ( since $\psi_a^{in}$ and $\psi_p^{in}$ being centered on the past
trajectories $x=\pm t$ have no overlap for $t\to -\infty$). In the future
it
behaves as a classical pair each member of which carries unit charge. This
is due to the remarkable formula
\begin{equation}
\lim_{t \to + \infty}[-{1 \over \a^* \b}  \psi^{in *}_a
\psi^{in*}_p]=
 \psi^{out }_p  \psi^{out *}_p + \psi^{out }_a  \psi^{out *}_a
\end {equation}
 obtainable
from the Bogoljubov transformation between in and out modes eq.
(\ref{apoviii})
plus the fact that the overlap of $\psi_a^{out}$ and $\psi_p^{out}$
vanishes in the distant future).

Figures IIIa--d display the real and imaginary part of $j_\mu(x)$ with the
$\psi$'s constructed from minimal wave packets ( eq. (\ref{avi}) with
$\sigma =
\sigma_{min}$). Unlike the pictures Figs Ia--c and Fig. II there is now no
advantage in
displaying pictures for $m/a ={\cal O}(1)$ since the normalisation factor
in eq. (\ref{bi})
brings the produced current up to order of magnitude unity (recall we are
working with
"conditional" amplitudes). We have chosen $m/a=9$. The WKB
approximation is then almost always valid, therefore the tunneling
interpretation
"de rigueur'' (as well as the whole theoretical framework). We now discuss
some interesting
features of these drawings.

In Section 2 we constructed packets, say $\psi^{in}_p(x)$, whose
branches are of fixed width $(=E^{-1/2})$ straddling the
corresponding classical trajectories. It may also be shown that for
$m/a>>1$,
in the tunneling region $\v x \v < a^{-1}$, the tunneling bridge also has
width
${\cal O}( E^{-1/2})$. This situation however changes radically when one
studies products such as $\psi_a^{in *}(x) \psi_p^{in *}(x)$ or derivatives
thereof such as in $j_\mu(x)$. Here the bridge thickens in time to give a
symetric space time tunneling region of dimensions $a^{-1}\times a^{-1}$.
This
is understood both physically and mathematically as follows.
Physically it has been
explained in \cite{bps} that one needs the space separation of a pair in a
vacuum fluctuation to be  ${\cal O}(a^{-1})$ in order for the negative
electric
energy ($=-E \v \Delta x \v$) to overcome the rest mass threshold.
Alternatively the virtual particles must be accelerated in a time interval
$\Delta t = {\cal O}(a^{-1})$ to pick up the energy necessary to overcome
the
threshold. Mathematically, the problem may be posed in either of the gauges
discussed in this paper. In space gauge, tunneling is in the region between
turning points ($\v \Delta x \v =2/a$) whereas in time gauge backscattering
occurs during a time interval  $\v \Delta t \v = {\cal O}(a^{-1})$. We also
may point
out that when WKB approximation applies to the tunneling region, there is a
euclidean classical path $x^2 + (Im t)^2 = a^{-2}$ which is used to get the
tunneling action. Since this is the result of a steepest descent
calculation wherein a contour has been distorted so as to give
imaginary values of $t$, it is not unexpected that as a function of
$x$ and ${ \rm Re} t$, the relevant production region is spread throughout
the circle of radius $a^{-1}$.

It should be noted that the description of production in terms of minimal
packets is a very precise representation of the physics in localized terms.
This is because these minimal packets serve as an exellent starting point
for the rigorous construction of complete orthogonal basis functions for
the
quantization of the field $\phi(x)$\cite{meyer}. More precisely, by
complete set we mean
the set necessary and sufficient to describe the modes which lead to pair
production in the space time region, ${\cal R}$, for which $E$ is non
vanishing  $-{L \over 2} \leq x \leq + {L \over 2}$, $0 \leq t \leq T$.
Such
modes are those that scatter off potentials whose centers lie within this
region. As shown in \cite{pb}, their number is $E L T / 2 \pi$. See end of
Appendix 1 for further discussion of the physics expressed.

Since the minimal wave packet has width $E^{-1/2}$ about the classical
orbit
in space-time, we see that if the packets are separated one from the other
by ${\cal O}(E^{-1/2})$, the number of them that can be fitted into ${\cal
R}$ is
${\cal O}(ELT)$. Thus with some tinkering on their size and shape
they will constitute a complete set, in the above sense, and orthogonal
because they are non overlapping.

One then comes upon the physical picture of production of quanta of minimal
size ${\cal O}(E^{-1/2})$ where the production zone is within a space-time
cell of size ${\cal O}(a^{-1} \times a^{-1}) $. For $m/a >> 1$, one may
think of
production as a set of shots emerging from cells. The number of packets
which contribute to a given cell is $ [ (a^{-1} \times a^{-1}) /E ] = m/a$.
We have displayed in Figs IIIa--d the production of a single pair of
these quanta.

Another noteworthy feature is the existence of oscillations in time that
occur within the
``circle" of production i.e. bounded by radius $a^{-1}$ wherein the
particles are still
virtual. In Section 1 it was stated that we were getting ``inside the
golden rule", and indeed
this is what we are now seeing inside the circle of production. To put
these oscillations into
evidence in this region it is most simple to examine the modes in temporal
gauge [eq.
(\ref{av})]. The function $D_{i \mu^2-1/2} \left [ e^{3 i \pi/4}
\sqrt{2}(t+ k)\right]$
is the solution of
\begin{equation}
\bigr[\partial^2_t +(Et+k)^2+m^2\bigl]D_{i \mu^2-1/2} \left [ e^{3 i \pi/4}
\sqrt{2}
(t+ k)\right]=0 \label{eqtg}
\end {equation}
whereupon it is seen that for small $t$ ($\v t \v <<m/E=a^{-1}$) the modes
oscillate
with frequency $\omega_k=\sqrt{k^2+m^2}$. For $t\leq a^{-1}$, one sees from
eq. (\ref{eqtg})
that the frequency of oscillation is still bounded by ${\cal O}(m^{-1})$.
These oscillations of
the modes are reflected in the oscillation of $j_{\mu}$ formed from wave
packets. For $x=0$
the frequency is then $m/(\pi\/ \sqrt{1+\sigma^4})$ obtainable from eq.
(\ref{avi}).
In Fig. IV
we have plotted ${\rm Re}[ \rho ]$ for $\sigma = \sqrt{2}$ whereupon one
sees the corresponding
decrease in the frequency. The picture emerges, then, that near the origin
the modes are
those of free particles, plane wave of momentum $k$. The pair creation
begins as a vacuum
fluctuation of free particles which subsquently upon separation is
converted into a real
pair. The oscillations are thus the manifestation the ``time--energy"
uncertainty relation
(in the sense of Fourier transform) during the phenomenon of creation. They
disappear upon
completion of the creation act. It is remarkable that the quantum number
$k$ gradually
changes its significance from a momentum to the time which dates the origin
of the creation
of the given pair. We call attention to the asymmetry in space and time
even though the
production region is symmetric. The uncertainty principle implies
ocillations in time and not
in space, in preparation for particles which propagate causally along the
forward light cone.
We also mention that the amplitude of these oscillation varies in space
time, approximatively
according to the law $\exp E(a^{-2}-t^{-2}-x^{-2})$.

Some of us conjecture that the decrease of the oscillations as
the width of the wave packets increases beyond its minimal value, is due to
the
averaging of the detector over the production of more than one
quantum.

Clearly this extreme sensitivity to the wave packet construction
necessitates
a very careful analysis of the detector system which is outside the
immediate
interest of this paper. Suffice it to say that we have produced a
description
of the source which is responsible for the back reaction onto the $E$ field
\underline{as} a pair is produced. It oscillates, is complex, and is
sensitive to the
resolving power of the detector, say a pair of coincidence counters of
varying
aperture. Outside the production region one's classical expectations are
fulfilled.

Since the title of this paper is Quantum Back Reaction it is incumbent upon
us
to integrate $\nabla . E = \rho$ to get the electric field due to the pair.
The
result is displayed in Fig. V.

A fundamental remark is now in order. At bottom what we have done is to
calculate a non
diagonal matrix element of $E$. From the fundamental point of view $E$ is
an operator
independently of whether we use gauss's law as a calculational tool to
evaluate its matrix
elements. Therefore there must exist a construction, called the wave
function of $E$ (better,
functional of $E(x)$). What we are doing is taking the very first steps to
construct it.

In consequence of the above considerations we believe that the back
reaction of a black hole due to emission of Hawking quanta will lead to
complex metrics in response to which the movement of the material source
of the black hole will display very specific quantum effects which will be
encoded in the
wave function of the metric. What the incidence of this will be on the
unitary issue
remains to be seen.

\section{{\bf Appendix 1:} Quantization in an $E$ field}

We first summarize quantization of a complex scalar field
in the gauge
$A_0 = -Ex$, $A_x=0$ in one spatial dimension. The Klein Gordon equation
is
\begin{equation}
({\cal D}_\mu {\cal D}^\mu  + m^2 )\phi =0 \label{apo}
\end{equation}
with ${\cal D}_\mu= \partial_\mu + i A_\mu$ the covariant derivative (the
electric charge $e$ has been englobed in the definition of $A_\mu$). It
posses
modes  of the form
\begin{equation}
\vf_\o (t,x) = C e^{i \o t} \chi_\o (x)\label{apoi}
\end{equation}
where $C$ is chosen to norm $\vf$ to unit incident flux
( $  \int\! dx \  i \vert C \vert^2 \left[ \vf_\o^* ({\cal D}_0 \vf_\o) -
({\cal D}_0^* \vf_\o^*) \vf_\o \right] =1$).
$\chi_\o (x)$ obeys the equation
\begin{equation}
\left [ { \partial^2 \over \partial x ^2}
+ E^2 (x-\o/E)^2 \right]  \chi_\o (x) = m^2 \chi_\o (x)\ .\label{apoii}
\end{equation}
Upon dividing by $-2m$ one comes upon a Schr\"odinger equation
describing tunneling of a particle of mass $m$ in an upside
down oscillator potential centered at $x=\o/E$ whose curvature
is $-E^2/2m$. The energy of the particle is $-m/2$ so that its
classical turning points are at $(x-\o/E)=\pm a^{-1}$ where $a$
is the acceleration ($=E/m$). In this effective Schr\"odinger
problem an incoming particle (in a wave packet) gives rise to
reflected and transmitted waves of flux $R$ and $T$
respectively and for unit incident flux we have the unitarity
condition
\begin{equation}
\v R \v^2 + \v T \v^2 =1\ .\label{apoiii}
\end{equation}
Since these two branches accelerate in opposite directions we
surmise that in the KG context they correspond to particles of
opposite charge and that the reflected and incident wave have
the same charge.

This identification is borne out by analysis of the motion of
wave packets of the modes $\vf_\o (t,x)$, that is linear
combinations of the form $\int d\o f(\o-\o_0) \vf_\o(t,x)$
(rather than the usual sum over energy eigenfunctions of the
Schr\"odinger equation).
 For example if $E$
points to the right and if an incoming packet
from the right carries unit positive charge one establishes
that the transmitted wave having amplitude $\b$ carries negative
charge (with flux $- \v \b \v^2$) and the reflected wave with
amplitude $\a$ carries positive charge  (with flux $- \v \a
\v^2$). Charge conservation yields
\begin{equation}
\v \a \v^2 - \v \b \v^2 =1\ .\label{apoiv}
\end{equation}

As explained in refs \cite{bps,pb}, the relation between eq. (\ref{apoiii})
and eq.
(\ref{apoiv}) is established through the swap of incident and reflected
waves
necessitated by the movements of the wave packets. Thus one
obtains eq. (\ref{apoiv}) from eq. (\ref{apoiii})  by dividing the latter
by $\v R \v^{-2}$
with $\a = 1/R$, $\b = T/R$. In this way $\b$ is identified
with the amplitude for pair production.

The modes $\vf_\o (t,x)$ satisfying the above initial
conditions (incident flux in the direction of $E$) provide
basis functions for the in-quantization scheme since they
correspond to the propagation of a single particle in the
past and a particle plus a pair in the future. Its parity
conjugate obtained by $(x-\o/E) \rightarrow - (x-\o/E)$ then
corresponds to the presence of an antiparticle in the past ,
once more yielding an additional pair in the future.
Introducing labels $p$, $a$ for particle (antiparticle) and
following the standard convention for Whittaker functions \cite{ww}
 we thus have the in-basis functions
\begin{eqnarray}
&\vf^{in}_{\o ,p}(x,t) &= { e^{-\pi \mu^2/4} \over (2 E)^{1/4} }
e^{i \o t}
D_{i \mu^2-1/2} \left [ e^{-3 i \pi/4} \sqrt{2 E}
(x-\o/E)\right]  \nonumber\\
&\vf^{in}_{\o ,a}(x,t) &=\vf^{in}_{-\o ,p}(-x,t)  \label{apov}
\end{eqnarray}
where $\mu^2 = m^2/2E$.
The second quantized field is then written, following the
usual rules
\begin{equation}
\vf = \sum_\o\ a^{in}_\o \vf^{in}_{\o ,p}
+ b^{in\ +}_\o \vf^{in *}_{\o ,a}\  .\label{apovi}
\end{equation}

Since the in-basis functions contain a particle plus a pair in
the future they are not useful to describe quantization in
terms of single quanta in the future (i.e. those which would
be registerd in counters). Their time reversed versions are
clearly  the right set and these are given simply by
\begin{eqnarray}
&\vf^{out}_{\o ,p}(x,t) &=\vf^{in *}_{\o ,p}(x,-t)  \nonumber\\
&\vf^{out}_{\o ,a}(x,t) &=\vf^{in *}_{-\o ,p}(-x,-t)\ .\label{apovii}
\end{eqnarray}

Since for each $\o$ the set $\vf^{in}_{\o ,p}$ and
$\vf^{in}_{\o ,a}$ is complete one may express the
$\vf^{out}_\o$'s as linear combinations of
$\vf^{in}_\o$'s. One finds \cite{bps}

\begin{eqnarray}
&\vf^{out}_{\o ,p} &= \a \vf^{in}_{\o ,p}
+ \b \vf^{in *}_{\o ,a}  \nonumber\\
&\vf^{out}_{\o ,a}&= \b \vf^{in *}_{\o ,p}
+ \a \vf^{in}_{\o ,a} \label{apoviii}
\end{eqnarray}
where
\begin{equation}
\a= {\sqrt{2 \pi}\ e^{-i\pi/4}e^{-\pi \mu^2/2} \over \Gamma(1/2 +
i \mu^2)} \qquad \b = e^{i\pi/2}e^{-\pi \mu^2}\label{apoix}
\end{equation}
Inserting eq. (\ref{apoviii}) into the expansion
\begin{equation}
\vf = \sum_\o\ a^{out}_\o \vf^{out}_{\o ,p}
+ b^{out\ +}_\o \vf^{out *}_{\o ,a} .\label{apox}
\end{equation}
one finds the Bogoljubov transformation
\begin{eqnarray}
&a^{out}_\o &= \a^* a^{in}_\o - \b^* b^{in +}_\o   \nonumber\\
&b^{out}_\o &= \a^* b^{in}_\o - \b^* a^{in +}_\o \label{apoxi}
\end{eqnarray}

In these terms one may calculate various matrix elements of
interest. For example the amplitude to find no out particles in
the future is \cite{schwing}
\begin{eqnarray}
 \v_{out}\la O\v O\ra_{in}\v^2= \prod_\o {1 \over \v\a\v^2}
&= exp -\! \sum_\o ln ( 1 + \v \b \v ^2 )   \nonumber\\
&= exp - {LTE \over 2 \pi} ln ( 1 + \v \b \v ^2 ) \label{apoxii}
\end{eqnarray}
where $LT$ is the volume of the spacetime box over which $E$ is
nonvanishing. See the end of this appendix for the proof and discussion of
$\sum_\o = (2 \pi )^{-1} L T E$.

We now sketch the corresponding analysis in the gauge
$A_x=Et$, $A_0=0$. It is in this gauge that we have carried out
our computations. Here the modes are of the form
$C e^{i k x} \xi_k(t)$ where $\xi_k(t)$ obeys the same type of
equation as $\chi_\o(x)$. This is obtained by replacing $x$ by
$t$ and $\o$ by $k$. The term $m^2 \chi_\o$ on the r.h.s. of
eq. (\ref{apoii}) becomes $-m^2 \xi_k$. Thus the effective Schr\"odinger
equation is the same but for the sign of the energy
($+m/2$ rather than
$-m/2$). Tunneling does not exist in these modes, rather it is replaced by
back-scattering in time. A packet moving forward in time in
the distant past gives rise to a transmitted wave propagating
forward in time and a reflected wave moving backward in time.
Following the analytic procedures of \cite{ww} one readily  confirms
that the amplitude of this backward wave is equal to the
amplitude of the transmitted wave in the corresponding
tunneling problem. Furthermore, analysis of wave packets
confirms one's expectations that the r\^ole of the transmitted
wave in $x$ gauge is played by that of the reflected wave in
$t$ gauge. Therefore up to a phase the coefficients $\a$ and
$\b$ remain the same and the whole previous analysis of
quantization in $x$ gauge is applicable as such. One simply
changes to the basis functions $e^{i k x} \xi_k(t)$ which in
terms of Whittaker functions are
\begin{equation}
\vf^{in}_{k ,p}(t,x) = { e^{-\pi \mu^2/4} \over (2 E)^{1/4} }
e^{i k x}
D_{i \mu^2-1/2} \left [ e^{3 i \pi/4} \sqrt{2 E}
(t+ k/E)\right]\ .\label{apoxiii}
\end{equation}
With this choice of function and using the same definition for the
antiparticle functions  and
the out-basis functions as in eq. (\ref{apov}) and eq. (\ref{apovii}) the
$\a$ and $\b$
coefficients are equal in both gauges and given by eq. (\ref{apoix}).

We close this appendix with a discussion of the number of modes \cite{pb}
($=\sum_\o$ in spatial gauge or $\sum_k$ in temporal gauge). We shall work
in
temporal gauge so that $\sum_k = (2 \pi )^{-1} L \int dk$. To calculate
$\int
d k$ it suffices to remark that the centers of the potentials $[ -
(t-k/E)]^2$
must occur within the region $0 \leq t \leq T$ i.e. while the $E$ field is
switched on. Thus $\int dk = E \int_0^T dt = ET$ so that $\sum_k = ELT/2
\pi$.

The above derivation highlights an interesting feature which is rather
novel
when one treats production as we do in terms of modes (better packets).
Normally one would pose the problem in terms of an $E$ field which depends
explicitly on time corresponding to switch on and off. The solutions for
modes would then be a difficult problem, but the point is that this
difficulty is irrelevant. It suffices to handle these modes which are
affected by the field during time $T$ and volume $L$ i.e. those for which
the
centers of the effective potential lie within $LT$ (up to edge effects
wherein the centers lie within $a^{-1}$ of the edge). For these relevant
modes our wave packets are good approximations to those which one would
have
with an explicit time dependent $E$ i.e. the true physical modes. The
others
have been falsly represented but they are irrelevant. When we speak of
vacuum
we mean vacuum with respect to the relevant modes presented in this paper.
the vacuum of irrelevant modes would have to be rewritten in terms of the
``true'' physical modes i.e. those which do not produce pairs within the
region $LT$. Since these are left out anyway we need not bother with them.
In
a word, we are replacing the space-time dependence of $E$ by following the
space-time dependence of modes (packets) and including only those that
would
have been affected by the true $E$. In this way the dynamical problem is
replaced by a static one supplemented by the proper counting procedure.

\section{{\bf Appendix 2:} Gaussian wave packets }

In this appendix we derive the second equality in eq. (\ref{avi}). To
prepare
for the integral over  $k$  we first obtain a slightly more general
integral representation of $\vf^{in}_{k,p}$ than eq. (\ref{av}) by applying
Cauchy's theorem to the contour $(0, \infty, \infty e^{i {\rm arg} \lambda}
,0)$
\begin{eqnarray}
 &\vf_{k, p}^{in} (x,t) &=  {{e^{-\pi \mu^2/4}}
\over {(2 E)^{1/4}}}
e^{i k x}
D_{i \mu^2-1/2} \left [ e^{3 i \pi/4} \sqrt{2 E}
(t+ k/E)\right]   \nonumber\\
& \ &=
{ e^{-\pi \mu^2/4} \over (2 )^{1/4} \Gamma(1/2-i\mu^2)}
\lambda^{-i \mu^2 +1/2}
e^{i k x} e^{i(k+t)^2/2}    \nonumber\\
& \ & \qquad \times \int_0^{+\infty} du\
e^{(1-i)(k+t)\lambda u - \lambda^2 u^2/2}
u^{-i\mu^2 - 1/2}\label{apti}
\end{eqnarray}
where $\v {\rm arg} \lambda \v < \pi /4$. A gaussian wave packet centered
on
$x_0=0$, $t_0=0$ is
\begin{eqnarray}
 &\psi^{in}_{p}(x,t) &= {e^{-\pi \mu^2/4} \over (2 )^{1/4}
\Gamma(1/2-i\mu^2)} \lambda^{-i \mu^2 +1/2}   \nonumber\\
 & \ & \qquad \times \int_{- \infty}^{+ \infty}\!\! dk\!
{e^{-{k^2 \over 2 \s^2}} \over \sqrt{2 \pi} \s}
\int_0^{+\infty}\!\! du\
e^{i k x +i{ (k+t)^2 \over 2} + (1-i)(k+t)\lambda u -
{\lambda^2 u^2 \over 2 }}
u^{-i\mu^2 - 1/2}   \nonumber\\
& \ &= { e^{-\pi \mu^2/4} \over (2 )^{1/4}
\Gamma(1/2-i\mu^2)} \lambda^{-i \mu^2 +1/2}
{1 \over \Sigma \s} e^{i { t^2 \over 2}}
e^{-{(t+x)^2 \over 2 \Sigma^2}}
  \nonumber\\
& \ & \qquad \times \int_0^{+\infty} \! du
e^{-\lambda^2  { 1 + i \s^2 \over 1 - i \s^2} { u^2 \over 2} }
e^{\lambda u (1-i) ( t + i (t+x) \Sigma^{-2} )}
u^{-i\mu^2 - 1/2} \label{aptii}
\end{eqnarray}
where $\Sigma^2 = {1-i \s^2 \over \s^2}$ and $\v {\rm arg} \Sigma \v <
\pi/4$.
In order that the permutation of the integrals carried out above be
meaningfull $\lambda$ must be taken such that $Re \lambda^2 > 0$ and
$Re ( \lambda^2  { 1 + i \s^2 \over 1 - i \s^2} )> 0$. The integral
over $u$ in eq. (\ref{aptii}) is of the same form as eq. (\ref{apti}) ,
therefore a
representation of a Whittaker function. By identification one obtains the
second line of eq. (\ref{avi}) in the text.

We note that the asymptotic behavior discussed in eq. (\ref{aiii}) et seq.
can be
recoverd by using the exact wave packets and the asymptotic expansion
for Whittaker functions \cite {ww}.

\section{{\bf Appendix 3:} Post--selection and weak measurment}

 In quantum mechanics one usually starts with an ensemble of
identically prepared systems, subjects them to different experiments and
analyzes
the probability distribution of the results. In such a situation
one deals with a ``pre--selected" ensemble: the preparation of the ensemble
took place before the experiments to which it was subjected. A ``pre-- and
post~--~selected" ensemble involves a supplementary step, a final
measurement.
According to the result of this final measurement the original ensemble is
split into
subensembles; each of these subensembles is called a pre and post selected
ensemble.
In each of these subensembles the distribution of the results of the
intermediate
measurements is different from that in the original, pre--selected only,
ensemble, and
depends on both the pre-- and the post--selection. As shown in \cite{aharo}
the
analysis of these distributions reveals both surprising physical effects
and
interserting features of quantum mechanics.

The most surprising effects occur when, between the pre-- and
post--selection, the
system interacts only weakly with test particles. Consider the following
scenario in which a variable A is ``weakly" measured. A system
prepared at time $t_1$ in the state $\v\p_1\ra$ evolves
according to a hamiltonian $H_0$ until a time $t_0$ when it interacts with
a
test particle (which can be viewed as the pointer of a measuring device).
The interaction hamiltonian is
\begin{equation}
H_{int}=\delta (t-t_0)Ap,\label{wmi}
\end{equation}
where $A$ is a variable of the system and $p$ is the canonical momentum of
the
test particle, conjugate to a position variable $q$. The above interaction
hamiltonian is the one used by von Neumann to model the standard
measurement of A. Indeed, during the interaction the effect of $H_0$, the
proper hamiltonian of the system can be neglected, and the time evolution
is
\begin{equation}
U(\tp, \tm)=\int_{\tm}^{\tp}e^{-i H_{int}(t)}dt=e^{-iAp}.\label{wmii}
\end{equation}
If the state of the system at $\tm$ is an eigenstate of $A$, say
$\v A=a\ra$, and the initial position of the pointer  precisely
defined, say in the state $\v q=0\ra$, the effect of the interaction would
be
\begin{equation}
U(\tp, \tm)\v A=a\ra\v q=0\ra=\v A=a\ra\v q=a\ra,\label{wmiii}
\end{equation}
that is, the pointer is moved from $q=0$ to $q=a$, showing the value of
$A$.
Furthermore, if at $\tm$ the system is in a superposition of eigenstates of
$A$, say $\sum_i c_i\v A=a_i\ra$, the system will get correlated with the
pointer
\begin{equation}
U(\tp, \tm)\sum_i c_i\v A=a_i\ra\v q=0\ra=\sum_i c_i\v A=a_i\ra\v q=a_i\ra.
\label{wmiv}
\end{equation}
When ``reading" the pointer we obtain different values $q=a_i$ with
probability $\v c_i\v$, as prescribed by the postulates of quantum
mechanics.

Suppose however, as opposite to von Neumann, that the initial position of
the
pointer is not accurately defined, say it is represented by a broad
gaussian
$\exp(-q^2/\Delta^2)$. As a result the measurement is imprecise
(even if one reads very precisely the final position of the pointer, one
still
does not know how much the pointer moved, as one does not know its initial
position exactly). On the other hand the system is only slightly disturbed
by the measurement. Indeed, the effective magnitude of the interaction
depends on the values of $p$ in eq. (\ref{wmii}), which in this case are
essentially
bounded by $1/\Delta$, supposed to be small. Alternatively $H_{int}$ may be
multiplied by a small coupling constant.
After this ``weak
measurement" the system evolves again according to $H_0$ until
a final time $t_2$ when a final measurement takes place and the system is
found to be in a state $\v\p_2\ra$. What is the final state of the
measuring
device corresponding to this sequence of events? Putting all this together
we
get
\begin{equation}
\Phi_{fin}=\la\p_2\v U(t_2,\tp)e^{-iAp} U(\tm, t_1)\v\p_1\ra
e^{-{{q^2}\over{\Delta^2}}},\label{wmv}
\end{equation}
where $\Phi_{fin}$ is the final state of the pointer. Since $\Delta$ is
large we can use a first order approximation to obtain
\begin{eqnarray}
&\Phi_{fin}&\approx\la\p_2\v U(t_2,\tp)(1-iAp) U(\tm,
t_1)\v\p_1\ra e^{-{{q^2}\over{\Delta^2}}}  \nonumber\\
&  &=\la\p_2\v U(t_2,\tp) U(\tm,
t_1)\v\p_1\ra (1-iA_wp)e^{-{{q^2}\over{\Delta^2}}},\label{wmvi}
\end{eqnarray}
where
\begin{equation}
A_w={{\la\p_2\v U(t_2,\tp)A U(\tm, t_1)\v\p_1\ra}\over
{\la\p_2\v U(t_2,\tp)U(\tm, t_1)\v\p_1\ra}}.\label{wmvii}
\end{equation}
The important point here is that one has added an increment to the detector
wave function which owing to the complexity of $A_w$ leads to unexpected
phase
dependent effects. One may see this in a picturesque way by
reexponentiation to
obtain
\begin{eqnarray}
&\Phi_{fin}&\approx\la\p_2\v U(t_2,\tp)U(\tm, t_1)\v\p_1\ra e^{-iA_wp}
e^{-{{q^2}\over{\Delta^2}}}\nonumber \\
& \ &={\rm Cte}\ e^{-{{(q-A_w)^2}\over{\Delta^2}}}
\nonumber \\
& \ &\approx {\rm Cte}\ e^{-{{(q-Re A_w)^2}\over{\Delta^2}}}e^{iqIm Aw}\ .
\label{wmix}
\end{eqnarray}
bringing into evidence the real and imaginary parts of $A_w$. Therefore
 the test particle which weakly interacted with the pre-- and
post--selected
system, behaves as if the variable A of the system has the complex value
$A_w$.
The real part of $A_w$ causes a shift in the position of the test particle
while
the imaginary part of $A_w$ produces a shift in its momentum.

The above discussion was done in Schr\"odinger representation. In
Heisenberg
representation $A_w$ becomes
\begin{equation}
A_w(t_0)={{\/ _{out}\! \la\p_2 \v A(t_0)\v\p_1 \ra _{in}}\over
{\/ _{out}\!\la\p_2 \v\p_1 \ra}_{in}},\label{wmx}
\end{equation}
where $A_w(t_0)$ is the Heisenberg operator $A$ evaluated at time $t_0$ and
where $\v\p_1 \ra _{in}$ and $\v\p_1 \ra _{out}$ represent incoming and
outgoing states
respectively.

\section{{\bf Appendix 4:} A decomposition of $G_F(x,x^\prime)$ }

We prove eq. (\ref{biii}) by availing ourselves of the identity
\begin{equation}
a_k^{out +} b_{k^\prime}^{out +} \v 0\ra_{out} =
{1 \over \a^{* 2} } a_k^{in +} b_{k^\prime}^{in +} \v 0\ra_{out}
-{ \b \over \a^*} \delta ( k - k^\prime) \v 0\ra_{out}\label{apfi}
\end{equation}
which follows from the definition of the Bogoljubov transformation eq.
(\ref{apoxi})
$a_k^{in +} = \a^* a_k^{out *} + \b b_k^{out}$ and similarly for $b_k^{in
+}$.
Use the definition of $\v 1_p 1_a \ra_{out}$ eq. (\ref{bii}) together with
eq. (\ref{apfi})
and $\phi$ expanded in the in-basis to give the required relation
\begin{equation}
{ \ _{out}\! \la 1_p 1_a \v
\phi^*(x) \phi (x^\prime) \v 0\ra_{in} \over
\ _{out}\! \la 1_p 1_a \v 0\ra_{in} }
=
-{ 1 \over \a \b^*} \psi^{in *}_{p}(x)
\psi^{in *}_{a}(x^\prime)
+ { \ _{out}\! \la 0 \v
\phi^*(x) \phi (x^\prime) \v 0\ra_{in} \over
\ _{out}\! \la 0 \v 0\ra_{in} }\ .\label{apfii}
\end{equation}

To prepare for Appendix 5 we derive the expression of
the in-out propagator as
a sum over modes. One first obtains
\begin{equation}
{ \ _{out}\!  \la 0 \v
\phi^*(x) \phi (x^\prime) \v 0\ra_{in} \over
\ _{out}\! \la 0 \v 0\ra_{in} } = \sum_{k, k^\prime}
\vf_{a,k}^{out}(x) \vf_{a,k^\prime}^{in *}(x^\prime)
{ \ _{out}\! \la 0 \v
a^{out}_k a^{in +}_{k^\prime} \v 0\ra_{in} \over
\ _{out}\! \la 0 \v 0\ra_{in} }\label{apfiii}
\end{equation}
whereupon insertion of the expression for $a^{out}_k$ given above yields
\begin{eqnarray}
&{ \ _{out}\! \la 0 \v
\phi^*(x) \phi (x^\prime) \v 0\ra_{in} \over
\ _{out}\! \la 0 \v 0\ra_{in} } &=
\sum_k {1 \over \a}
\vf_{a,k}^{out}(x) \vf_{a,k}^{in *}(x^\prime)   \nonumber\\
& \ &=  \ _{in}\! \la 0 \v
\phi^*(x) \phi (x^\prime) \v 0\ra_{in} + \sum_k {\b \over \a}
\vf_{p,k}^{in *}(x) \vf_{a,k}^{in *}(x^\prime) \label{apfiv}
\end{eqnarray}
where $\ _{in}\!\! \la 0 \v
\phi^*(x) \phi (x^\prime) \v 0\ra_{in} =
\sum_k \vf_{a,k}^{in}(x) \vf_{a,k}^{in *}(x^\prime)$ is the propagator in
in-vacuum.

\section{{\bf Appendix 5:} Hadamard's substraction }

In order to compute finite quantities quadratic in the
field it is necessary to regularize the matrix elements
of $\phi^*(x_0) \phi(x_0)$. The regularisation scheme we
adopt is to subtract the Hadamard function, i.e. we
subtract the solution of the field equations that most
ressembles the Minkowskian propagator in a neighborhood
of a point $x_0$\cite{mpb,mas}. The subtraction is universal insofar
as it is independent of the matrix element to be
calculated (it is constructed using only the field and
its derivatives at $x_0$). Therefore when taking the
difference of matrix elements of $\phi^* \phi$ (as discussed in
Section 3 to subtract operationaly the noise term appearing
in eq. (\ref{biii} )) the Hadamard function cancels out. To express this
universality of the subtraction we introduce the following
expression for the renormalized product of two field
operators
\begin{equation}
\phi^*(x) \phi(x_0) \v_{ren} =
\phi^*(x) \phi(x_0) - G_H (x, x_0) I\label{apfivei}
\end{equation}
where $G_H (x, x_0)$ is the Hadamard function and it is
understood that the limit $x \rightarrow x_0$ is to be taken
at the end of any computation. $I$ is the identity matrix in
Hilbert space so that the Hadamard function cancels in the difference
of matrix elements discussed in Section 3.

Upon renormalizing according to eq. (\ref{apfivei}) the matrix elements
appearing in Section 3 always contain a finite term plus a
noise term equal to $G_F(x,x_0) - G_H(x,x_0)$ (where $G_F$ is the
in-out propagator). In Section 3 the finite term was
discussed at length. In this appendix we consider the noise
term. To this end we first obtain closed forms for $G_H$ and
$G$.

To construct the Hadamard function it is of conceptual interest to
carry through the calculation for a general space-time
varying electromagnetic field $F_{\mu \nu}$. The informed
reader will then recognize the more familiar analogous
construction used in the presence of gravity. One first
chooses a gauge such that the wave equation coincides at
$x_0$ with the free field equation, i.e. by taking
$A^\mu(x_0)=0$ and $\partial_\mu A^\mu (x)=0$; in addition
one requires that the derivatives of $A^\mu$ at $x_0$ depend
only on $F_{\mu \nu}$ and its derivatives at $x_0$
(this choice of gauge -- Poincar\'e's gauge -- is the analog of the riemann
normal
coordinates in curved space time).
Explicitly this is obtained by taking
\begin{equation}
A_\mu (x^\b) = \int_0^1 t (x^\a - x^\a_0 ) F_{\a \mu}
(x^\b_0 + t (x^\b - x^\b_0 ))\/ dt \label{apfiveii}
\end{equation}
(For a constant electric field this gauge is $A_0=-Ex/2$, $A_x
= Et/2$). The Hadamard function is then fixed by requiring
that its short distance behavior in this gauge be that of a
free field
\begin{equation}
G_H(x,x_0) \mathrel{\mathop{\approx}_{\lambda \rightarrow 0}} {i\over
2\pi}\bigl(\ln
{\sqrt{m^2\lambda}\over2} +\gamma\bigr) \label{apfiveiii}
\end{equation}
where $\lambda = (t -t_0)^2 - (x - x_0)^2$.

The Hadamard function can be expressed for finite $\lambda $ as
\begin{equation}
G_H(x,x_0) = e^{-i E (xt-x_0 t_0)/2} H_{\mu^2} (E \lambda) \label{apfiveiv}
\end{equation}
where $H_{\mu^2} (z)$ (with $\mu ^2 =m/2a$) is a solution of
\begin{equation}
 \bigl( z{d^2\over dz^2}+{d\over dz}+{z\over 16}+2\mu^2) H_{\mu^2} (z)
=0\label{apfivev}
\end{equation}
therefore of the form
\begin{equation}
 H_{\mu^2} (z) =e^{-i{E\lambda\over 4}}\big[a\/U({1\over 2}
+i\mu^2,1,i{E\lambda\over 2}) + b\/M({1\over 2}
+i\mu^2,1,i{E\lambda\over 2})\label{apfivevi}
\end{equation}
where $U$ and $M$ are confluent hypergeometric functions (Kummers'
functions).
 In order to satisfycondition eq. (\ref{apfiveiii}) the constants $a$ and
$b$ have to be equal to
\begin{eqnarray}
&a &=-{i\over 2\pi} \Gamma({1\over 2}+i\mu^2)  \nonumber\\
&b &= {i\over 2\pi}\bigl[\ln \mu^2-{i \pi \over 2}-\psi({1\over
2}+i\mu^2)\bigr]
\label{apfivevii}
\end{eqnarray}

A closed form for the in-out propagator can be obtained from
its representation as an integral over Schwinger fifth time
derived in refs \cite{schwing,stephens,bps}. The equivalence of this
representation
with the sum over modes given in Appendix 4 has been proven
in \cite{parentani}. Using formula [1IV.8] of ref. \cite{ast} it is
straightforward to verify that in the gauge eq. (\ref{apfiveii})
\begin{equation}
G_F(x,x_0) = e^{-i E(xt-x_0 t_0)/2} \Delta_{\mu^2} (E
\lambda)\label{apfiveviii} \end{equation}
\begin{equation}
\Delta_{\mu^2} (z) =-{i\over 2\pi}\Gamma({1\over 2}+i\mu^2)U({1\over 2}
+i\mu^2,1,i{E\lambda\over 2})\label{apfiveix}
\end{equation}

We are now in a position to discuss the physical content of
the noise term. The charge and current carried by this term
are zero as is seen by acting on $G_F(x,x_0) -
G_H(x,x_0)$ with the differential operator $i ( { \cal
D}_{\mu x_0} - { \cal
D}_{\mu x}^* )$. For short distances we have in Poincar\'e's gauge
\begin {equation}
G_F(x,x_0)-G_H(x,x_0)\approx _{\l \rightarrow 0} -b\/\exp
{-iE(xt-x_0t_0)/2}\bigl(1+({1\over 2}+i\mu
^2)i{E\l \over 2}+{\cal O}(\l ^2)\bigr)\quad.
\end {equation}
As the charge and the current are obtained from the action of a first order
differential operator,
only the first phase factor of this expression will contribute in the
coincidence limit and we
obtain~:
\begin {eqnarray}
&\rho_{in,out} &=\lim_{(t,x)\rightarrow
(t_0,x_0)}i\bigl[(\partial_{t_0}-iEx_0/2)- (\partial_{t}-iEx/2)\bigr]exp
{-iE(xt-x_0t_0)/2} \nonumber \\
& \ &=0\qquad ,\nonumber \\
&{in,out} &=\lim_{(t,x)\rightarrow
(t_0,x_0)}i\bigl[(\partial_{x_0}+iEt_0/2)- (\partial_{t}+iEx/2)\bigr]exp
{-iE(xt-x_0t_0)/2} \qquad ,\nonumber \\
& \ &=0\qquad .
\end{eqnarray}
This result is confirmed by calculating the
current carried by the in-out propagator when it is expressed
as a sum over modes whereupon it is seen to vanish identically.
As this confirms that when considering the charge the noise term in eq.
(\ref{biii})
vanishes we prove it below.

The vanishing of the charge is most conveniently
exhibited by working in temporal gauge wherein it occurs mode by mode.
Explicitly using eq. (\ref{apfiv}) one obtains
\begin{eqnarray}
&\rho_{in,out}(k) &= \vf^{out}_{p,k} (x,t)(i \lr{\cal D}_0)
\vf^{in *}_{p, k}(x,t) +
 \vf^{out}_{a,k} (x,t)(i\lr {\cal D}_0)^* \vf^{in *}_{a, k}(x,t)
\nonumber\\
& \ &= e^{ikx}\chi^*(-t-k) ( i \lr{\partial}_t) e^{-ikx} \chi^*(t+k)
\nonumber \\
& \ & \qquad + e^{-ikx}\chi^*(-t-k) (- i \lr{\partial}_t) e^{ikx}
\chi^*(t+k)  \nonumber\\
& \ &=0\label{apfivex}
\end{eqnarray}
where in passing from the first to the second line we have
expressed all the modes in terms of $\chi (t+k)$ as in eqs
(\ref{apov},\ref{apovii}).
To exhibit the vanishing of $j_x$ it is convenient to pass
to the spatial gauge whereupon $j_x$ vanishes mode by mode for
the particles and antiparticles seperatly
\begin{eqnarray}
&j_{in,out}(\o) &=
\vf_{p,\o}^{out} (i \lr{\cal D}_x) \vf_{p,\o}^{in *}  \nonumber\\
& \ &= e^{i \o t} \chi^*(x-\o) ( i\lr \partial_x ) e^{-i \o t} \chi^*(x-\o)

\nonumber\\
& \ &=0\label{apfivexi}
\end{eqnarray}
since the second line is the Wronskian of two identical functions and thus
is zero.

Another interesting property of the noise term to calculate is the
expectation
value of $\phi^*(x_0)\phi(x_0)$ (i.e. the coincidence point of
$G_F(x,x_0)-G_H(x,x_0)$) since the imaginary part of $G_F(x_0,x_0)$ is
related to
the rate $\Gamma$ of pair creation per unit time and unit length  by
\begin{equation}
\Gamma T L = {\rm Im}\int \! d^2x \int_{m^2}^{+\infty}\! dm^2 ( G_F(x,x)
)\label{apfivexii}
\end{equation}
(where $LT$ is the space time dimension of the region wherein
$E\neq 0$). It is therefore a natural propriety of the formalism ( easily
verified from the explicit expression for $G_H$) that in the
coincidence limit the imaginary part of $G_F(x_0,x_0)$ is unchanged by the
renormalisation.  The real part of $G_F(x_0,x_0)-G_H(x_0,x_0)$ is related
in
similar manner to the renormalised energy of the vacuum.

It is instructive to rewrite this last equation as a sum over the members
of
a complete orthonormal set of localized wave packets
(see Appendix 4)
\begin{equation}
\Gamma T L = {\rm Im}\int \! d^2x \sum_k \int_{m^2}^{+\infty}\! dm^2 \left[
\v \b
\v^2  {1 \over \b^* \a} \phi^{in *}_{a,k}(x) \phi^{in *}_{p,k}(x)
\right]\label{apfivexiii}
\end{equation}
whereupon it is seen that it is expressed as the sum
over all possible pairs of the imaginary part
of $\int dm^2 \phi^* \phi$ if that
pair was created (i.e. the same term as  eq. (\ref{bv} ) times the
probability of
creation of that pair ($=\v \b \v^2$).  Since for a
fixed pair the imaginary part of $\phi^* \phi$ is localized withing the
corresponding cell,
the volume dependence in eq. (\ref{apfivexii}) can be expressed as a sum
over all
cells times the probablity that a pair be created in that cell.

\centerline{\bf{Figure  Captions}}

{\bf Figs. Ia, Ib, Ic} \\
Representation by equally spaced contour lines of the absolute value of
wave packets
$\v \psi^{in}_p \v$ (eq. (\ref{avi}) with $\s = \s_{min}$) centered on the
origin of space-time.
 We have illustrated $\v \psi^{in}_p \v$ for $m/a =1/4$ (Fig. Ia) as
typical for $m/a
< 1$ and for $m/a =9$ (Fig.Ib) as typical for $m/a >>1$. In the latter case

the created antiparticle does not appear since its amplitude is to small to
see.
Fig. Ic is a drawing of $\ln \v \psi^{in}_p \v$ for $m/a =9$ which permits
display of
the created anitparticle. In all figures the length unit is the inverse
acceleration
$a^{-1}$.

{\bf Fig. II} \\
Representation by equally spaced contour lines of the charge density $\rho
(t,x)$ carried by the
wave packet drawn in Fig. Ia . The created antiparticle (on the left)
carries
negative charge.

{\bf Figs. IIIa, IIIb, IIIc, IIId} \\
Figs IIIa--d are plots of the charge and current due to a pair created in a
minimal wave packet
for $m/a =9$. Fig. IIIa is ${\rm Re }[ \rho (t,x)]$; Fig.IIIb is ${\rm Im
}[ \rho
(t,x)]$; Fig. IIIc is ${\rm Re }[ j (t,x)]$; Fig. IIId is ${\rm Im }[ j
(t,x)]$. These plots
do not
describe the true amplitude of the oscillations as if $ j_\mu $ is greater
than
$+1$ or less than $-1$ it is drawn in white or black irrespectively to
 its true value.ratherblack and withe simply designate the signs while gray
shows the
transition between them. As mentionned in the text the variation of the
amplitude as a
function of $x$ and $t$ is very large.It is seen upon inspection of Figs
IIIa--~d that the
region of production is confined to the expected circular cell of radius
$\simeq {\cal
O}(a^{-1})$. Outside this region ${\rm Re }[ j_\mu (t,x)]$ is due to
particles moving on
classical orbits and ${\rm Im }[ j_\mu (t,x)] \simeq 0$. Within the
tunneling region ${\rm Im
}[ j_\mu (t,x)]$ is of the same order of magnitude as ${\rm Re }[ j_\mu
(t,x)]$ and both
oscillate.

{\bf Fig. IV} \\
The real part of the charge density $\rho(t,x)$ due to a created pair in
a gaussian wave packet with $m/a =9$, $\s = \sqrt{2} \s_{min}$ is ploted
using the same conventions
as Fig. III. Compared to it  the amplitude
and frequency of the oscillations have decreased.

{\bf Fig. V} \\
A picture of the real part of the electric field obtained by integrating
Gauss's law
$\nabla . E = \rho$ when the charge density is that of the pair drawn in
Fig.~IIIa. The electric
field oscillates in the region where the pair is created whereas for late
times when the particles
are on mass shell the back reaction electric field becomes constant between
the particles.

\end{document}